\documentstyle[12pt,amssymbols]{article}
\input math_macros.tex

\begin{document}
\begin{titlepage}
\begin{center}
December 14, 1994     \hfill    LBL-36346 \\

\vskip .5in

{\large \bf Is Mental Process Non-Computable?}
\footnote{This work was supported by the Director, Office of Energy
Research, Office of High Energy and Nuclear Physics, Division of High
Energy Physics of the U.S. Department of Energy under Contract
DE-AC03-76SF00098.}
\vskip .50in

\vskip .5in
Henry P. Stapp\\
{\em Lawrence Berkeley Laboratory\\
      University of California\\
    Berkeley, California 94720}
\end{center}

\vskip .5in

\begin{abstract}
It has recently been claimed that certain aspects of mental processing cannot
be simulated by computers, even in principle. The argument is examined and a
lacuna is identified.

\end{abstract}
\end{titlepage}
\renewcommand{\thepage}{\roman{page}}
\setcounter{page}{2}
\mbox{ }

\vskip 1in

\begin{center}
{\bf Disclaimer}
\end{center}

\vskip .2in

\begin{scriptsize}
\begin{quotation}
This document was prepared as an account of work sponsored by the United
States Government. While this document is believed to contain correct
 information, neither the United States Government nor any agency
thereof, nor The Regents of the University of California, nor any of their
employees, makes any warranty, express or implied, or assumes any legal
liability or responsibility for the accuracy, completeness, or usefulness
of any information, apparatus, product, or process disclosed, or represents
that its use would not infringe privately owned rights.  Reference herein
to any specific commercial products process, or service by its trade name,
trademark, manufacturer, or otherwise, does not necessarily constitute or
imply its endorsement, recommendation, or favoring by the United States
Government or any agency thereof, or The Regents of the University of
California.  The views and opinions of authors expressed herein do not
necessarily state or reflect those of the United States Government or any
agency thereof or The Regents of the University of California and shall
not be used for advertising or product endorsement purposes.
\end{quotation}
\end{scriptsize}

\vskip 2in

\begin{center}
\begin{small}
{\it Lawrence Berkeley Laboratory is an equal opportunity employer.}
\end{small}
\end{center}

\newpage
\renewcommand{\thepage}{\arabic{page}}
\setcounter{page}{1}

\noindent{\bf 1. Introduction}\\
Roger Penrose has recently published an argument$^1$ that seeks to establish
that mathematicians, when they come to know mathematical truths, {\it cannot}
in all cases be relying solely on processes that can be adequately
simulated by idealized computers. Within the framework of science
this is a startling claim, for contemporary mainstream scientific thought
holds that mental processing insofar as it leads to overt behaviour is an
aspect of physical processes happening mainly in the brain, and that
these processes are governed by the mathematical laws of classical and quantum
physics, and hence should be able to be simulated to arbitrary accuracy, at
least in principle, by computers, provided no practical limitations whatsoever
are imposed. Penrose's argument seeks to refute this. Moreover, the argument
is claimed to be close to rigorous. Thus it is claimed, in effect, that
almost rigorous argumentation is able to demolish some tenets of mainstream
scientific thought, and to demonstrate that fundamentally new ideas are
therefore required. This conclusion, if valid, would be a breakthrough of major
importance in science.

\noindent{\bf 2. Penrose's Argument}

1. Let $C_q(n)$, for $q$ ranging over some infinite set $R_q$, be a listing of
all computational processes that depend on one natural-number argument $n$.
For each pair $(q,n)$ the computational process $C_q(n)$ either stops, or never
stops. (Example. $C_{7}(n)$ might be: Find the smallest integer $N\geq 0$ that
is not
a sum of $n$ numbers each of which is a square of a natural number,
$0, 1, 2, 3, ...$.  For $n\geq 4$ no systematic search for $N$ will ever stop,
according to a theorem due to Lagrange)

2. Proceed by {\it reductio ad absurdum}: Assume that if, for some pair
$(q,n)$, we can know that $C_q(n)$ can never stop then we can know this only by
means of some reasoning processes that, because it is the reflection of an
underlying brain process, can be assumed to be a computational process that
yields an answer, and thus stops (because it can be programmed to stop when it
yields an answer). Thus the {\it reductio ad absurdum}
assumption is that if, for some pair $(q,n)$, we can know that $C_q(n)$ can
never stop then there must be some computational process $A(q,n)$ such that:

`$A(q,n)$ stops' implies `$C_q(n)$ can never stop'.

3. If $A(q,n)$ is defined for every pair $(q,n)$ (see below) then $A(n,n)$ is a
computational process that depends on one argument, $n$.
Then there must be an index $k$ such that:
$$
A(n,n)=C_k(n).
$$

4. Therefore, according to the assumption of line 2,

`$C_k(k)$ stops' implies `$C_k(k)$ can never stop'.

5. Therefore,

$C_k(k)$ can never stop.

6. But (by line 3) $C_k(k)=A(k,k)$, and hence (by line 5)  `$A(k,k)$ can never
stop'.

7. Thus we have found out that `$C_k(k)$ can never stop',
yet the knowledge that `$C_k(k)$ can never stop' is not entailed by line 2.

8. We conclude that the $A(k,k)$ occurring in line 2 for the case
$(q,n)=(k,k)$ is not unique: there must be an $A^1(k,k)\neq A(k,k)$ whose
stopping entails that `$C_k(k)$ never stops'. (Penrose specifies that the
stopping of $A(q,n)$ is merely a sufficient condition for $C(q,n)$ never to
stop, not a necessary and sufficient one. Hence there might be several
different processes $A^m(k,k)$ any one of which could serve as the $A(k,k)$
in line 2.)

9. If there were only a finite number of processes $A^m(k,k)$ such that the
stopping of any one of them would allow us to know that $C_k(k)$ can never stop
then one could define the $A(k,k)$ in line 2 to be the process that stops if
and only if any one of these $A^m(k,k)$'s stops. Then one would get the desired
contradiction: We would know (by line 5) that `$C_k(k)$ can never stop', yet
(by line 6) that the unique computational process whose
stopping would (according to line 2) allows us to know this fact can never
stop.

Penrose$^1$ has argued that all of the $A^m(q,n)$ whose stoppings can allow us
to know that $C_(q,n)$ can never stop, as specified in line 2, can indeed be
amalgamated into one single $A(q,n)$. In this case, the assumption in line 2
becomes: for any pair $(p,n)$, if `` `$C_p(n)$ can never
stop' is knowable'' then ``$A(q,n)$ stops'', and, conversely, for any pair
$(p,n)$, if ``$A(p,n)$ stops'' then `` ` $C(p,n)$ can never stop' is knowable''
Thus we have the equivalence: for every pair $(p,n)$,

``$A(p,n)$ stops'' iff `` `$C(p,n)$ never stops' is {\it knowable}''
\vspace{.5in}

Since whatever is knowable is (presumably) true the argument can then proceed
as indicated above, with a contradiction appearing after line 6. However,
there is a question about line 3, to which we now turn.

\noindent{\bf 3. Indices and Arguments}

Let us consider a set of processes $S_q(n)$, where $q$ ranges over an infinite
set $R_q$. It is useful to make a
distinction between an {\it index}, represented by a subscript, and an
{\it argument}, represented by a variable enclosed by parentheses. The
dependence on an {\it argument} is supposed to be one in which some {\it
finitely-stated} rule covers the infinite set of values that the argument
(for example, the natural number $n$) is allowed to take on, whereas the
dependence on an {\it index} is supposed to be one that is expressed by means
of a case-by-case listing of the infinite set of individual cases. In the
former case,  the various possible values of the argument are elements of a
coherent mathematical structure (e.g., the set of natural numbers), which makes
it possible for one finitely-stated rule to cover the infinite number of
possible values of the argument. But in the latter case the full identity of
the index is specified, say, by its shape: the symbol is identified exclusively
by an intrinsic identifying characteristic, not by means of the logical
connections of this symbol to the other ones. Thus one could use
 *, !, ?, [, ...
for these intrinsically characterized symbols, instead of 0, 1, 2, 3, ... ,  to
indicate their lack of logical relatedness to one another.

The processes $C_q(n)$ were introduced by Penrose by {\it listing} all of the
different computational processes $C(n)$  that are functions of the single
(natural-number) argument $n$:
$$
C_0(n), C_1(n), C_2(n), C_3(n), C_4(n), C_5(n), \ldots .
$$
This way of introducing the set of $C_q(n)$ might suggest that $q$ is an
index, and hence that, in my notation, it is properly written as a subscript,
which is how Penrose writes it.

In this case, where the set of all possible $C_q(n)$  is indexed by the set of
subscripts $q$, where q ranges over a set $R_q$ of pure symbols,
the set of processes $A(q,n)$ should be written rather
as $A_q(n)$: the set of symbols $q$ would be merely a set of indices, each of
which has an identity, but no logical relationship (apart from `different
from')
to the others. Hence no rule apart from direct case-by-case listing is possible
for specifying the dependence on $q$.

If one were to adhere to this point of view, that the symbols
$q$ are merely indices, then Penrose's argument would collapse. For, it would
make no sense to say that a pure symbol, say $*$, is equal to some natural
number $n$. If one were, in spite of this logical point, simply to set up a
convention whereby the pure symbols were represented by natural numbers in
some haphazard way then one could not expect to derive anything useful. One
could then, to be sure, formally consider the set of processes $A_n(n)$,
as $n$ runs over the set of natural numbers. But this set could not coincide,
for some $k$, with the set of $C_k(n)$'s, for all $n$, because the
dependence of $A_n(n)$ upon the subscript $n$ is
{\it not} of the argument type, whereas for each value of $q$ the dependence
of $C_q(n)$ upon $n$ {\it is} of the argument type, by definition. Thus a key
step in Penrose's argument, namely line 3, would fail.

Penrose certainly recognized that he would not obtain a valid argument if the
symbol $q$ were an index-type of variable: he specified that $q$ must be
regarded as an argument-type of variable, but did so without ever writing down
$C(q,n)$. Once one writes $C(q,n)$ instead of
$C_q(n)$ a question immediately arises: How can one confirm that there is, in
fact, a computational process $C(q,n)$ that depends on two {\it arguments},
and has the property that, as $q$ runs over the natural numbers, the
process $C(q,n)$ runs over the complete set of processes that are functions of
the other argument $n$? Specifically, if the set of all computable processes
of one (natural number) argument $n$ is the set of $C_p(n)$, with $p$ running
over its range $R_p$, then how does one construct a {\it finitely described}
computational process
$C(q,n)$ that acts on two (natural-number) {\it arguments} $q$ and $n$, such
that for every $p$ in $R_p$ there is a natural number $q_p$ such that
$C(q_p,n)= C_p(n)$.

Penrose$^1$ answers this question satisfactorily. He considers
a G\"{o}del-type of construction whereby one imagines that there is some rule
whereby {\it the sequence of mathematical symbols} that expresses the form of
each computational process $C_p(n)$ is transcribed into some corresponding
natural number $q_p$, in such a way that $ C_p(n)=C(q_p,n)$ for each $p$ in
$R_p$.

Let it be granted, therefore, that $C_q(n)$ can, in my notation, be replaced
by $C(q,n)$. Then the
computability assumption (that must be shown to lead to a contradiction)
asserts that for every
pair $(q,n)$ such that `` `$C(q,n)$ can never stop' is knowable'' there is a
computational process $A_{q,n}$ that stops and is such that:\\
`` `$A_{q,n}$ stops' implies `$C(q,n)$ can never stop' ''.\\ This condition is
the (reductio ad absurdum) assertion that the only way that one can know
that `$C(q,n)$ can never stop' is by means of a mental process that can be
represented by a computational process$^1$.

To complete the proof described in section 2 one must show that set of
processes $A_{q,n}$ can be represented in the form $A(q,n)$; i.e., that the
dependence of $A_{q,n}$ on the two {\it indices} $q$ and $n$ can be
represented by an {\it argument-type} of dependence, not merely by an
index-type dependence. An index-type of dependence is all that one is
allowed to assume, {\it ab initio}, without begging the question.

A proof that this $A_{q,n}$ has the form $A(q,n)$ would allow one to justify
line 3 of the proof. However, the assumption that there exists
a {\it fixed finitely stated rule} that maps the
arguments $(q,n)$ that identify any `process $C(q,n)$ that can be known never
to stop' '' onto ``the process $A$ by means of which it can be proved that
$C(q,n)$ never stops'' is a far more mind-boggling idea than the result that
is to be derived from this assumption. If it were true, it would mean that the
search for solutions of the various diverse and difficult individual problems
in number theory could in principle be avoided: there would exist a
{\it fixed finitely-stated rule} that maps the numbers that identify the
problem to be solved (if it can be solved) onto the very argument by means of
which it can be solved. The existence of such a general fixed finitely-stated
rule for solving all of the soluable problems in number theory
goes far beyond what can reasonably be expected.

What this means is that the assumption that $A_{q,n}$ can be written in the
form $A(q,n)$ (i.e., that the dependence of the process $A_{q,n}$ on the
variables $q$ and $n$ that identify $C(q,n)$ is a fixed finitely-stated rule)
begs the question: it must be proved, not assumed.

\noindent {\bf 4. G\"{o}delization}

One  might try  to  deal with  this  problem  by  exploiting  the deep  results
obtained by K.  G\"{o}del$^2$. In this  connection it  should be noted that the
assumption in line  2 goes far beyond  what was proved (in  this connection) by
G\"{o}del, who claimed (in terms of  the computer  formulation used by Penrose)
merely that

the set $K$  of $n$ such that  ``  `process $C(n,n)$ never stops'
is provable'' \\
is  characterized  by the   statement

 ``$C(k,n)$ stops'',\\
where $k$ is some well defined number  that is  explicitly constructable within
that formalism.

This diagonalized version of the assumption in line 2 is all that
is really needed for the proof. So there is  the possibility that a full
G\"{o}del-type argument might provide what is needed to complete the proof. But
then G\"{o}del's argument pertaining to what is {\it provable} on the basis of
certain mathematical rules known to mathematicians must be carried over to what
is {\it knowable} to human beings by virtue of hypothesized mechanical rules of
brain process. These latter rules act at the atomic level, and they can never
be
known to human beings in the same way that mathematical rules are known to
mathematicians: what is knowable to human beings rests on the coherency of what
they are aware of, not on their understanding of their own brain processess.

What Penrose  is trying to  refute is the  hypothesis that  what is knowable to
a human  being is determined  mechanically, in  terms of brain
activities  that are governed  by mechanical  rules. Since  what we can know is
presumeably a  mere surface  activity of a far more extensive brain activity,
it becomes  important to distinguish what  we know, or can know, from the more
extensive activity upon which it  rests. Within the computer framework that
Penrose is using, a conceivable model of the mind/brain could be this: the
brain  activity  is  represented by a   mechanical/computer  activity that {\it
stops} from  time to  time, and the output  represents the  conscious thought.
This output is then fed back into the  computer as the next input. The machine
is designed to  produce   outputs at a fairly  regular pace, and to terminate
any procedure that  does not  give  an output  reasonably quickly:  brains
must get answers out expeditiously if the organism is to survive.

In applying a G\"{o}del-type  argument to this  mind/brain system the
analog of the  mathematical rules in  G\"{o}del's work  would be the rules that
govern the activity of the brain. The conclusion of the G\"{o}del-type argument
(transcribed into the computer  language) would be that there must be an
allowed
brain  process $P$  that  can never  stop in  spite  of the  fact that  no
system
operating according to the rules by  which the brain operates could ever  reach
the conclusion that $P$ can never stop.

So the problem is: How can {\it we} reach this conclusion if no system
operating
according to the rules that govern the actions of our brains could ever reach
this conclusion?

Although the  hypothesized mechanical  rules that govern  brain action use some
elements of simple  arithmetic, there  is no need for them  to use  any process
that depends upon the use of the concepts ``for all $n$'', or  ``there exists
no
$n$'', or any  other  notion in which  is imbedded  the notion of  infinity.
The
simple   step-by-step  approximate  integration  of the    discretized forms of
differential equations of classical and quantum physics does not  encounter
any need to  answer infinite numbers
of  questions: the  questions it  encounters are of  the  finite  kind, such as
``what is  $1+1$ ?''  In fact,  every number  that  occurs in the  constructive
process of solving  these  finite-difference equations is  a finite number, and
these numbers, since they represent values that can occur in living brains, are
restricted  to certain  finite  domains. However,  this does not  mean that the
finite output statements  of these brains cannot  include the finite strings of
symbols that  are used by   mathematicians to  express  propositions  of number
theory that refer to infinite sets.

What happens  to G\"{o}del's proof  if one replaces  the mathematical rules
that are used in his argument by a strictly finitistic arithmetic that contains
no universal quantifiers such as ``for  all $n$ \ldots'', and that restricts
all
numbers to pre-specified finite sets.  The answer is that  the proof does not
get off the  ground, for it  rests heavily  on the concept of ``for all $n$''
and an unbounded domain for the natural numbers. Consequently, the
assertion that there exists a $k$ such that:

``$C(k,k)$ stops'' iff `` `$C(k,k)$ can never stop' is knowable''\\
cannot be proved within the finitistic type of model of the mind/brain
described above. So this attempt to supply the missing relation
$A_{q,n}=A(q,n)$ fails.

The finite-type computer $B$ that simulates the mechanical activity of the
human brain (and whose {\it outputs at stopping points} represent human
thoughts) can be imbedded in a computer $C$ whose rules of operation
included implementation of the concept `` for all $n$'', and to
which the G\"{o}del/Turing argument can be applied. Then a (super-human)
mind M that could comprehend both the rules of operation of $C$ and also the
logic of the G\"{o}del/Turing proof would be able to compute a value $k$
such that the following proposition $P_k$ is true:

``$C(k,k)$ stops'' iff `` `$C(k,k)$ can never stop' is {\it knowable}$_C$''

\noindent where {\it knowable}$_C$ means knowable by virtue of the {\it
outputs} of $C$. The mind $M$ that knows that $P_k$ is true can know also
that ``$X$ is {\it knowable}$_C$'' entails that ``$X$ is {\it true}'', and can
therefore conclude that ``$C(k,k)$ can never stop''. Thus $M$ can know
more than what is knowable$_C$. This is the analog of G\"{o}del's theorem, and
is not a contradiction. On the other hand, the human mathematician can know
only the output of $B$. He will be able to reason, on the basis of what the
hypothetical $M$ is
able to know, that there {\it exists some} $k$ (unknowable to human beings)
such that ``$C(k,k)$ can never stop''. However, it has not been proved that
the only way that he could know this is by virtue of the stopping of $C(k,k)$.
The stopping of $C(k,k)$ may be the unique process in $C$ whose stopping gives
the strong result that `` ` $C(k,k)$ can never stop' is knowable$_C$'', for the
particular value of $k$ that is specified by the G\"{o}del/Turing argument.
But no proof is offered that there can be no process in $B$ whose stopping
could establish the {\it far weaker} conclusion that ``{\it there exists a} $k$
such that
$P_k$ is true'', which is all that is known to the human mathematician. Indeed,
the human mathematician reasons on the basis of the general assumptions and
properties known to him, and these do not include any knowledge of the details
of the construction of $C$. He obtains from his reasonings conclusions that
do not refer to the specific details of the construction of $C$, and that are
therefore far weaker than the strong conclusion available to $M$. Penrose does
not show that there could be no process in $B$ whose stopping would yield this
far weaker conclusion. In the absence of such a demonstration no contradiction
is established, and hence the {\it reductio ad absurdum} argument fails to go
through.

{\bf References}\\
1. Roger Penrose, Shadows of the Mind: A Search for the Missing Science of
Consciousness, Oxford University Press, Oxford, 1994. Sect. 2.5\\
2. Kurt G\"{o}del, Monatshefte f\"{u}r Mathematic und Physik {\bf 38}, 173-198
(1931); and in {\it From Frege to G\"{o}del} ed. Jean van Heijennoort, Harvard
University Press, Cambridge MA (1976) pp. 596-617

\end{document}